\documentclass[12pt]{iopart}

\usepackage[dvips]{graphicx}

\begin{document}

\title{The KATRIN Pre-Spectrometer at reduced Filter Energy}

\author{M. Prall$^1$, P. Renschler$^2$, F. Gl\"uck$^{3,4}$,
A. Beglarian$^3$,\\
H. Bichsel$^5$,
L. Bornschein$^3$ ,
Z. Chaoui$^6$, 
G. Drexlin$^3$, \\
F. Fr\"ankle$^3$,
S. G\"orhardt$^3$,
S. Mertens$^3$, 
M. Steidl$^2$,  \\
Th. Th\"ummler$^3$,
S. W\"ustling$^3$,
C. Weinheimer$^1$, \\
S. Zadorozhny$^7$ }

\address{$^1$Universit\"at M\"unster, Institut f\"ur Kernphysik, Wilhelm-Klemm-Stra{\ss}e 9, \\
48149 M\"unster, Germany\\
$^2$Universit\"at Karlsruhe (TH), Institut f\"ur Experimentelle Kernphysik, \\ Postfach 6980, 76128 Karlsruhe, Germany \\
$^3$Karlsruher Institut f\"ur Technologie, Institut f\"ur Kernphysik, Postfach 3640, \\ 76021 Karlsruhe, Germany   \\
$^4$KFKI, RMKI, H-1525 Budapest, POB 49, Hungary \\
$^5$CENPA University of Washington P.O. Box 354290 Seattle, WA 98195-4290      \\
$^6$Laboratory of Optoelectronics and Devices, Faculty of Science,\\ University of Setif, Algeria\\
$^7$Institute for Nuclear Research of Russian Academy of Sciences, Moscow, Russia }

\ead{matthias.prall@uni-muenster.de}

\begin{abstract}
The {\bf KA}rlsruhe {\bf TRI}tium {\bf N}eutrino experiment, {\bf KATRIN}, will determine the mass of the electron neutrino with a sensitivity of \mbox{0.2 eV (90\% C.L.)} via a measurement of the $\beta$-spectrum of gaseous tritium near its endpoint of $E_0=18.57\;\textrm{keV}$. An ultra-low background of about $b=10$~mHz is among the requirements to reach this sensitivity. In the KATRIN main beam-line two spectrometers of MAC-E filter type are used in a tandem configuration. This setup, however, produces a Penning trap which could lead to increased background. 
We have performed test measurements showing that the filter energy of the pre-spectrometer can be reduced by several keV in order to diminish this trap. 
These measurements were analyzed with the help of a complex computer simulation, modeling multiple electron reflections both from the detector and the photoelectric electron source
used in our test setup. 
\end{abstract}


\maketitle

\section{The KATRIN Experiment}
\label{sec::intro}

\begin{figure}[!htbp]
\begin{center}
\includegraphics[width=\textwidth]{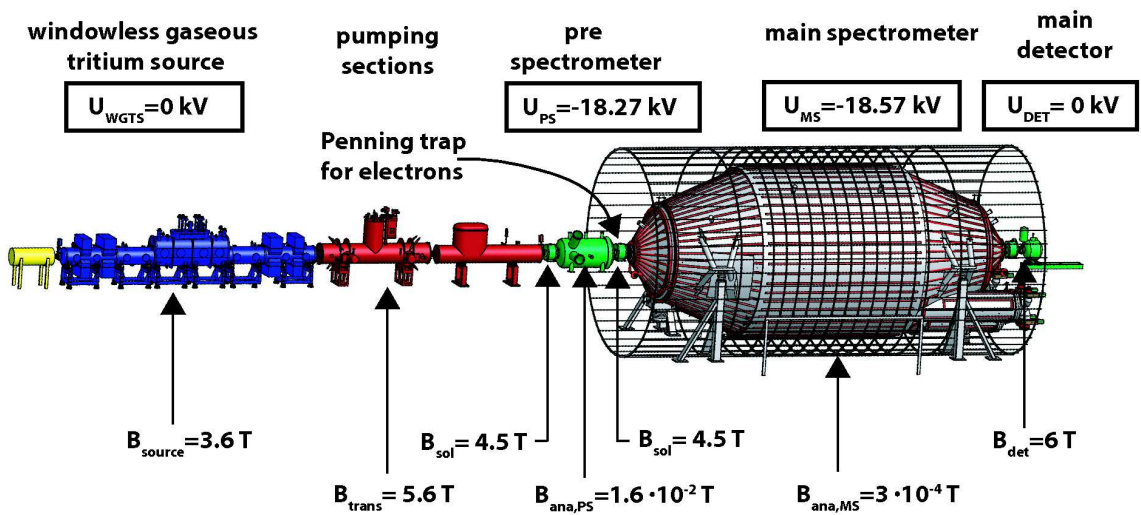}
\caption{Overview over the KATRIN experiment with the potentials and magnetic field strengths given in the KATRIN design 
report \cite{Ang04}. With these values, the setup contains a Penning trap for electrons between the pre- and main-spectrometer at $B_{\textrm{sol}}=4.5\;\textrm{T}$.
A change of the pre-spectrometer potential from $U_{\textrm{PS}}=-18.27\;\textrm{kV}$ by several kV towards smaller absolute values will diminish this trap.}
\label{fig::katrin_overview}
\end{center}
\end{figure}

The KATRIN experiment \cite{Ang04} will determine the mass of the electron
antineutrino\footnote{KATRIN will not be able to resolve the different neutrino mass eigenstates $m_{\textrm{i}}$, but will determine a weighted average $m^{\textrm{(eff)}}$ 
of the neutrino mass states $m_{\textrm{i}}$ according to their mixing $U_{\textrm{ei}}$ with the electron neutrino.}
 $m^{\textrm{(eff)}}_{\nu_e}=\sqrt{ \sum m_i^2 |U_{\textrm{ei}}|^2 }$ via a high precision measurement of the $\beta$-decay kinematics at the endpoint $E_{\textrm{0}}=18.57$~keV of the $\beta$-spectrum of tritium with a sensitivity of \mbox{0.2 eV}  \mbox{(90\% C.L.)}. The latest upper limits obtained with this model-independent method and the isotope tritium as $\beta$-emitter are from the experiments at \mbox{Mainz: $m^{(eff)}_{\nu_e}<2.3$~eV (95\% C.L.) \cite{Kra05}} and Troitsk: $m^{(eff)}_{\nu_e}<2.05$~eV (95\% C.L.) \cite{Ase03}. 
Figure \ref{fig::katrin_overview} shows a schematic overview over the 70 m long KATRIN setup: $\textrm{T}_2$-gas with an activity of $10^{11}$~Bq is recirculated in the so-called windowless gaseous tritium source (WGTS). 
 $\beta$-decay electrons are guided by a magnetic field towards the spectrometers. Both the pre- and main spectrometer are of the MAC-E filter type \cite{Pic92b}.

Electrons are guided through these spectrometers along the magnetic field lines and are decelerated by an electric filter potential $U_{\textrm{f}}$ in each of the spectrometers. 
$U_{\textrm{f}}$ is the high voltage with negative polarity applied to the spectrometer, or more exactly, to its inner electrode system \cite{Ang04}. In KATRIN, we use the independent filter potentials $U_{\textrm{f}}=U_{\textrm{PS}}$ for the pre- and $U_{\textrm{f}}=U_{\textrm{MS}}$ for the main-spectrometer. Only those electrons with an energy $E$ larger than the filter energy $qU_{\textrm{f}}$ (see Section \ref{sec::mac_e_principle}) are transmitted through a spectrometer and are reaccelerated to their original energy. Here $q=-e$ is the negative electron charge, and the filter energy $qU_{\textrm{f}}$ is the maximum potential energy of an electron in the spectrometers. The pre-spectrometer (PS) with an energy resolution of $\Delta E_{\textrm{PS}} \approx 100$~eV is the first filter for the $\beta$-decay electrons. It reduces the flux of $\beta$-decay electrons into the 
main-spectrometer (MS), lowering the rate of background electrons created in collisions with residual gas molecules \cite{Ang04}. Having an energy resolution of $\Delta E_{\textrm{MS}}=0.93$~eV, the MS scans the last 30 eV of the $\textrm{T}_2$ $\beta$-spectrum which contain the information on the neutrino mass. Finally, the electrons transmitted by the MS are counted by a 148 pixel PIN diode with an energy resolution of $\Delta E_{\textrm{det}}\approx 1$~keV.\\
\\
The motivation for our investigations is the following: Inside the MS, the $\beta$-decay electrons can start multi-step processes leading to free electrons. These can be accelerated towards the detector and
to energies around 18 keV by the MS potential. The energy resolution of KATRIN's detector is about $\Delta E_{\textrm{det}}\approx 1$~keV. Therefore, these electrons cannot be distinguished from signal electrons produced by the tritium $\beta$-decay. Thus, the background can rise above KATRIN requirement of $b=10$~mHz \cite{Ang04}. Therefore, the flux of $\beta$-decay electrons into the MS should be kept low.
The flux of $\beta$-electrons can be minimized by keeping the filter potentials of MS and PS relatively close (e.g. $U_{\textrm{PS}}=-18.27\;\textrm{kV}$ and  $U_{\textrm{MS}} \approx -18.57\;\textrm{kV}$). However, using the B-fields ($B_{\textrm{sol}}=4.5$~T, Fig. \ref{fig::katrin_overview}) mentioned in the KATRIN 
design report \cite{Ang04} the region between the two spectrometers is a Penning trap for electrons. By multi step processes this trap can lead to increased background as well \cite{Bec10}.
A reduction of the pre-spectrometer potential from $U_{\textrm{PS}}=-18.27\;\textrm{kV}$ by several kV towards smaller absolute values will diminish this trap. The optimum value of $U_{\textrm{PS}}$ which minimizes the background has to be determined experimentally. If $qU_{\textrm{PS}}$ is reduced by several keV however, the $\beta$-electrons with energies close to $E_{\textrm{0}}=18.57$~keV will retain a surplus energy $E_{\textrm{sur}}=E_{\textrm{0}}-qU_{\textrm{PS}}$ in the order of several keV inside the PS. Thus, electrons have higher speed and may no longer be guided by the magnetic field. This behaviour, leading to transmission losses, was already observed in the MAC-E filter of the Mainz Neutrino Mass Experiment \cite{Pic92a,Thu07}. For KATRIN, 100\% transmission above the PS filter energy $qU_{\textrm{PS}}$ is required. 
We present two main results in this publication:

\begin{enumerate}
	\item The requirement of 100\% transmission at reduced PS filter energy $qU_{\textrm{PS}}$ is fulfilled. If it should turn out that the Penning trap between the two spectrometers cannot be
suppressed by other means \cite{Bec10}, the PS filter energy can be reduced by many keV in order to overcome this problem.
	\item The KATRIN collaboration is able to model the electron transport and the electron backscattering at the detector with high precision in agreement with experimental data. This allows detailed investigations of the 
	experimental setup.
\end{enumerate}
This publication is organized as follows: In section \ref{sec::mac_e_principle} we review the operation principle of the MAC-E filter, Section \ref{sec::setup} presents our experimental setup; our measurements
and their analysis via custom simulation tools are presented in Sections \ref{sec::measurements} to \ref{sec::simulation}. Section \ref{sec::pre_spec_modes} discusses the background and adiabaticity
at reduced PS filter energy in detail. Finally, our findings are summarized in Section \ref{sec::conclusion_and_outlook}.

\section{The MAC-E Filter Technique}
\label{sec::mac_e_principle}

In this Section, we explain the principle of a MAC-E filter \cite{Pic92b} under standard conditions, 
i.e. when the filter energy $qU_{\textrm{f}}$ is close to (a few 10 eV)  the energy $E$ of the incoming electrons. \par

\begin{figure}[!htbp]
\begin{center}
\includegraphics[width=\textwidth]{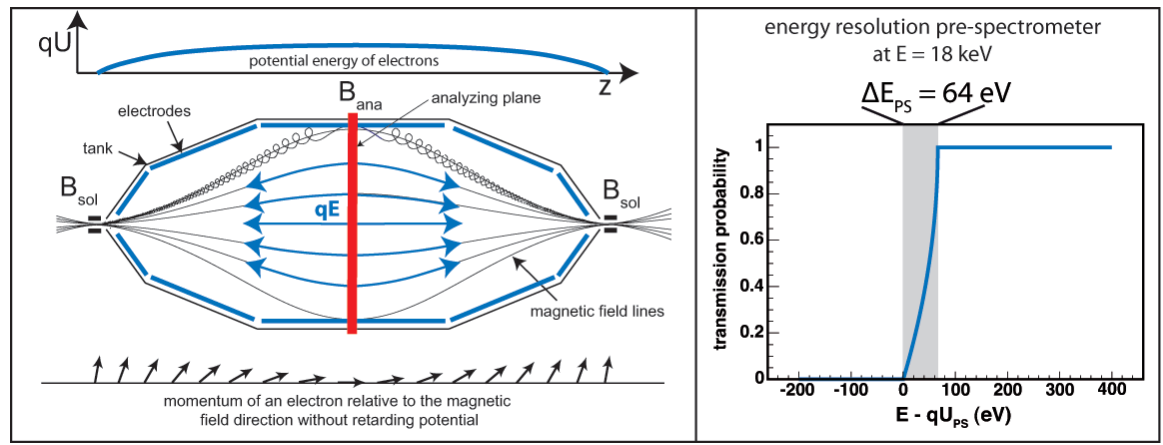}
\caption{Left: Working principle of a MAC-E filter. Right: Ideal transmission function of the pre-spectrometer for an isotropic source of monoenergetic electrons with $E_{\textrm{0}}=18.57$~keV, $B_{\textrm{ana,PS}}=0.016$~T and $B_{\textrm{sol}}=4.5$~T. The transmission probability of the pre-spectrometer (of a MAC-E filter) starts to become non-zero as soon as the energy $E$ of the incoming electron is larger than the filter energy $qU_{\textrm{PS}}$.
Both the tank and electrodes of the KATRIN pre- and main-spectrometer are on negative high voltage. Detailed explanations can be found in the main text.}
\label{fig::mac_e_principle}
\end{center}
\end{figure}
The principle of a MAC-E filter is illustrated in Fig. \ref{fig::mac_e_principle}: Two identical solenoids provide a guiding magnetic field. Ignoring drift motions which appear as higher order
corrections \cite{Leh64}, the $\beta$-electrons enter the MAC-E filter and follow the guiding magnetic field lines along helix-like trajectories resulting from the cyclotron motion.
This statement is true if the relative changes of the electric and magnetic field strength within a cyclotron length $l_{\textrm{cyc}}$ are small \cite{Leh64}:

\begin{equation}
\begin{array}{l}
\frac{|\Delta \vec{E|}}{|\vec{E}|}<<1 \;\;\;	  \textrm{and}   \;\;\;  \frac{|\Delta \vec{B}|}{|\vec{B}|}<<1	\\
\\
 \textrm{within one cyclotron length}  \;\;\; l_{\textrm{cyc}}:  \;\;\;
 l_{\textrm{cyc}} = 2 \pi \frac{v_{\parallel}}{\omega_{cyc}}  = 2 \pi \frac{\gamma m_{\textrm{e}}}{|q|B} v_{\parallel},
 \end{array}
\label{eq::cyclotron_length}
\end{equation}

Here, $v_{\parallel}$ is the electron velocity parallel to the guiding magnetic field line, $\omega_{cyc}$ the cyclotron frequency, $m_{\textrm{e}}$ the electron mass, $q=-e$ the negative electron charge and
$\gamma$ the relativistic factor.

If Eq. \ref{eq::cyclotron_length} holds, there is a conserved adiabatic invariant $\gamma\mu$, where $\mu$ denotes the orbital magnetic moment of the 
electron (see Section 12.5 in \cite{Jac99})

\begin{equation}
\mu = \frac{  \gamma+1  }{ 2\gamma }\cdot \frac{ E_{\perp} }{B  } = \frac{p_{\perp}^2}{2 m_{\textrm{e}} \gamma B }.
\end{equation}

$E_{\perp}=E_{\textrm{kin}}\cdot \sin^2 (\theta)$ is the fraction of the kinetic energy $E_{\textrm{kin}}$ which can be attributed to the motion around the guiding B-field line. $\theta$ is the polar angle 
between the guiding B-field line and the electron momentum vector. In the following, we will use the symbol $\varphi$ for the corresponding azimuthal angle. $E_{\parallel}=E_{\textrm{kin}}\cdot \cos^2(\theta)$ is the fraction of the kinetic energy connected to the forward motion of the electron. Only $E_{\parallel}$ is analyzed by the MAC-E filter. $p_{\perp}$ denotes the fraction of the electron momentum perpendicular to the guiding B-field line.

For KATRIN, the maximum electron energy is $E_{\textrm{0}}=18.57$~keV, thus one has $\gamma \leq 1.04$. Therefore, $\mu$ is a good approximation for the conserved quantity, especially when electrons are slowed down 
by the electric field in the MAC-E filter:

\begin{equation}
\mu \approx \frac{E_{\perp}}{B} = \frac{E_{\textrm{kin}}\cdot \sin^2{\theta}}{B} \approx const.
\label{eq::angle_formula}
\end{equation}

From Eq. \ref{eq::angle_formula} it is clear that the polar angle $\theta$ and $E_{\perp}$ are completely determined by the B-field and the kinetic energy of the electron. As B decreases towards the analyzing plane of the MAC-E filter, $E_{\perp}$ is minimized, providing the good energy resolution of the MAC-E filter. Electrons are guaranteed to be transmitted along the guiding B-field line if their initial energy is large enough to overcome the spectrometer potential. The transmission probability $T(E,qU_{\textrm{f}})$ of a MAC-E filter is derived from Eq. \ref{eq::angle_formula} by integrating over all electrons which fulfill $0<E_{\parallel}=(E-qU_{\textrm{f}})-E_{\perp}$ in the central plane of the MAC-E filter. These electrons start at ground potential.
Isotropically emitted electrons with starting energy $E$ and $\theta$ from $0^{\circ}$ to  $90^{\circ}$ have to be considered. Eq. \ref{fig::mac_e_principle} is used to transform between $E_{\perp}$ at the origin of the electrons with $B_{\textrm{s}}$ and the analyzing plane, with the minimum B-field strength $B_{\textrm{ana}}$:

\begin{equation}
\fl T(E,qU_{\textrm{f}}) =  1  -    \sqrt{      1-   \frac{E -qU_{\textrm{f}}}{E} \cdot  \frac{ B_{\textrm{s}}  }{ B_{\textrm{ana}}   }  }  \;\;\;\;\;\;\;\;   \textrm{for}	  \;\;\;\; \;\;\;\;   qU_{\textrm{f}}  \leq E \leq qU_{\textrm{f}} \cdot  \frac{  B_{\textrm{s}}  }{   B_{\textrm{s}} - B_{\textrm{ana}}  }   .
\label{eq::ideal_transfunct}
\end{equation}

Thus, the transmission probability $T(E,qU_{\textrm{f}})$ only depends on the magnetic field strengths, on the energy $E$ of the incoming electron and on $qU_{\textrm{f}}$, the filter energy. Below the interval specified in Eq. \ref{eq::ideal_transfunct}, the transmission probability is zero, above this interval it is unity (Fig. \ref{fig::mac_e_principle}). The transmission function describes an energy
high-pass filter, only electrons with an energy $E$ above the filter potential $qU_{\textrm{f}}$ are transmitted. The energy resolution of the MAC-E filter $\Delta E_{\textrm{f}}$ is equal to the maximum $E_{\perp}$ of an electron in its analyzing plane. $E_{\perp}$ in the analyzing plane is maximal if the polar angle $\theta$ is equal to $90^{\circ}$ in the entry-side magnet. The conservation of $\mu$ allows to compute 
the resolution $\Delta E_{\textrm{f}}$:

\begin{equation}
const. \approx \mu = \frac{E_{\perp}}{B} \;\;\;\; \Longrightarrow \;\;\;\; \Delta E_{\textrm{f}} = E \cdot \frac{B_{\textrm{ana}}}{B_{\textrm{s}}}.
\end{equation}

Considering electrons starting in the entry-side solenoid of the PS, one has to insert \mbox{$B_{\textrm{s}}=B_{\textrm{sol}}=4.5\;\textrm{T}$} and $B_{\textrm{ana,PS}}=0.016\;\textrm{T}$. 
The formula gives an energy resolution $\Delta E_{\textrm{PS}}=64\;\textrm{eV}$ if the PS is used with $E=18\;\textrm{keV}$ electrons (cf. right side of Fig. \ref{fig::mac_e_principle}). 
For a MAC-E filter and in the adiabatic approximation \cite{Leh64}, the Lorentz force

\begin{equation}
	\vec{F}=q\left(\vec{E} + \vec{v}\times \vec{B}\right).
\label{eq:lorentzForce}
\end{equation}
does not only result in a helix-like motion around the guiding magnetic field
line but also in an azimuthal magnetron drift around the spectrometer symmetry axis \cite{Pic92b}.

\begin{figure}[!ht]
\begin{minipage}[b]{8cm}
\centerline{\includegraphics[height=3.3cm]{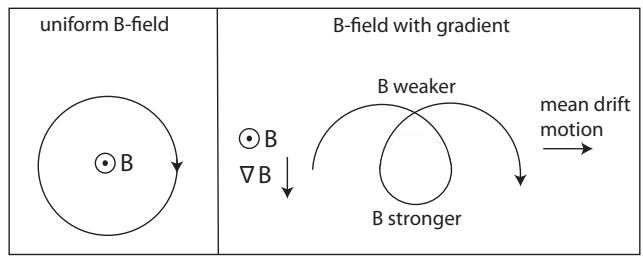}}
\end{minipage}
\hfill
\begin{minipage}[b]{8cm}
\setlength{\fboxsep}{0mm}
\caption{\textbf{Pictorial explanation of the Grad BxB-dift.} Left: The orbit of a charged particle in a uniform B-field. Right:
The orbit of a charged particle in a non-uniform B-field.}
\label{fig::explanation_brad_b_b_drift}
\end{minipage}
\end{figure}
Electrons which pass the MAC-E filter on its symmetry axis gyrate around the central magnetic field line in a helical cyclotron motion. Since the magnetic field is axially symmetric and the cyclotron radius is changing slowly, the electrons are in a quasi-constant magnetic field. For off-axis electrons, however, the magnetic field is asymmetric during a cyclotron motion. The radius of curvature of the electron trajectory is smaller in a stronger B-field.  Therefore, the gradient $\vec{\nabla} B$ results in an azimuthal $\vec{B} \times \vec{\nabla} B$ drift of the guiding center along a circle with constant magnetic field (Fig. \ref{fig::explanation_brad_b_b_drift}): 

\begin{equation}
\vec{v}_{\perp} =   \frac{1}{qB}    \left(  E_{\perp}   +   2\cdot  E_{\parallel}    \right)    \cdot \frac{ \vec{B} \times \vec{\nabla} B}{ B^2}.
\label{eq::drift_perp}
\end{equation}
The lower $qU_{\textrm{PS}}$ is in our investigations, the larger are $E_{\perp}$ and $E_{\parallel}$ and the azimuthal drift (\ref{eq::drift_perp}). The following section will 
show that our PS data can only be understood if the azimuthal drift (\ref{eq::drift_perp}) is taken into account.

\section{The Pre-Spectrometer Test Setup}
\label{sec::setup}

This section describes the experimental setup used to measure the PS transmission with $E=18$~keV electrons and a PS filter energy $qU_{\textrm{PS}}$ down to 1 keV, so that electrons retain
a surplus energy $E_{\textrm{sur}}=E - qU_{\textrm{PS}}$ of up to 17 keV in the PS. 

\begin{figure}[!htbp]
\begin{center}
\includegraphics[width=\textwidth]{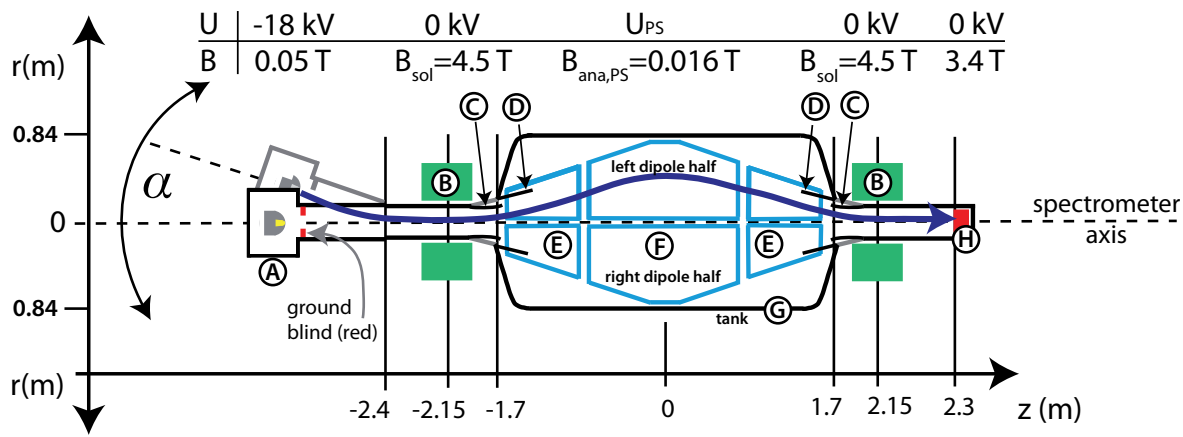}
\caption{The pre-spectrometer test setup: A) Photoelectric electron source (E-GUN), B) solenoid, C)  ground electrode, D) shielding electrode, E) full metal cone electrode, F) wire electrode, G) spectrometer 
vessel, H) 64-pixel silicon PIN diode. There is a longitudinal gap splitting electrodes E) and F) into a left and right half. For our measurements, the left and right dipole half were electrically connected and therefore
supplied with identical voltages. The dark blue arrow illustrates an electron trajectory.}
\label{fig::prespec_setup}
\end{center}
\end{figure}

The B-field of the PS test setup (Fig. \ref{fig::prespec_setup}) is generated by two solenoids (B). The inner electrode system of the PS tank (G) consists of four parts: The ground electrodes (C) define the potential at the entry and exit of the PS, the shielding electrodes (D) were introduced to avoid a Penning trap leading to background \cite{Fra10}, electrodes (E) are conical metal shields, the central part (F) is a wire electrode. The tank (G) and the shielding electrodes (D) are electrically connected. Both the tank and the electrodes are on negative high voltage. 
There is a longitudinal gap splitting electrodes (E) and (F) into a left and right half. For our measurements, the left and right dipole half were electrically connected and therefore supplied with identical voltages.
Previous measurements \cite{Fra10} with this setup and a pressure of $10^{-10}$~mbar inside the PS resulted in an average background rate of $17\pm 0.4$~mHz in the 
energy window from 15 to 21 keV over the whole detector. For these, the tank (G) and shielding electrode (D) were kept at $U_{\textrm{tank}}=-18\;\textrm{kV}$, the inner electrodes (E) and (F) were put on 
$U_{\textrm{electrode}}=-18.5\;\textrm{kV}$. As this configuration does not produce any background related to particles stored in Penning traps, the same potential difference
$U_{\textrm{electrode}} - U_{\textrm{tank}}=-0.5$~kV between the tank (G) and the electrode system (E,F) was used in our measurements. 
The potential inside the PS tank is a mixture of the electrode and tank potential, one has $U_{\textrm{PS}} = a\cdot U_{\textrm{tank}}+b \cdot U_{\textrm{electrode}}$. 
Yet, the constants are $a=0$ and $b=1$ in good approximation. $U_{\textrm{PS}}$ and $U_{\textrm{electrode}}$ never differ by more than a few tens of V.
This effect depends on the electron trajectory in the PS and is accounted for in our simulations (Section \ref{sec::simulation}). For these, we have computed the actual electric field inside the PS using the methods described in \cite{Glu11b,Glu11c}. In the following we do not distinguish between $qU_{\textrm{PS}}$ and $qU_{\textrm{electrode}}$ in the text as their difference is negligible at keV surplus energies.
In each of our measurement series, the tank voltage $U_{\textrm{tank}}$ was varied from -0.5~kV to about -17.5~kV, so that the PS filter energy  $qU_{\textrm{PS}}\approx qU_{\textrm{electrode}}$ varied from about 1~keV to about 18~keV. 

\begin{figure}[!htbp]
\begin{center}
\includegraphics[width=\textwidth]{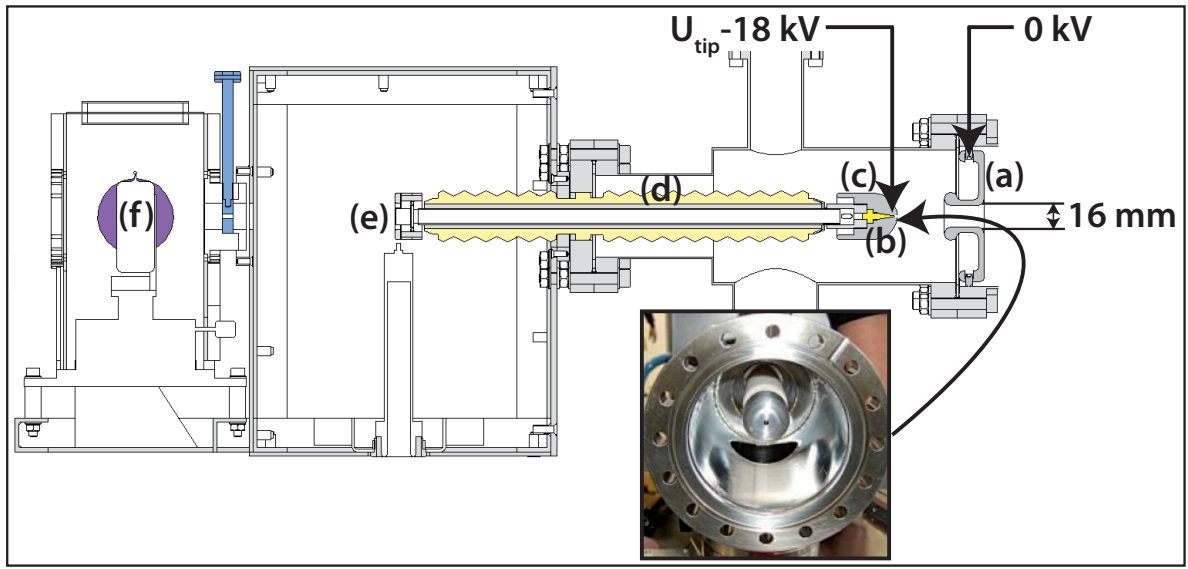}
\caption{The photoelectric electron source (E-GUN). a) ground blind (diameter of bore=16 mm), b) gold plated quartz tip at $U_{\textrm{tip}}=-18\;\textrm{kV}$ (yellow), c) metal housing (grey), d) insulator, e) sapphire window, f) deuterium UV lamp.
The inset shows the metal housing which includes the gold plated quartz tip. The ground blind was removed here.}
\label{fig::E-GUN}
\end{center}
\end{figure}

A photoelectric electron source (E-GUN) (Fig. \ref{fig::E-GUN}) mounted at the entry of the PS test setup (Fig. \ref{fig::prespec_setup}) was used to generate electrons with an energy of $E=18$~keV for the measurements:  A deuterium lamp (f) generates UV light with wavelengths in the range $185 \; \textrm{nm} < \lambda < 400 \; \textrm{nm}$ ($6.7 \; \textrm{eV} > h\nu > 3.1 \; \textrm{eV}$). The light shines through a sapphire window (e) and a hollow ceramic insulator (d). The UV photons finally produce free electrons via the photoelectric effect in a thin gold layer ($\approx 35\; \mu \textrm{g}/\textrm{cm}^2$) on a gold plated quartz tip (b) sitting in a metal housing (c).
The quartz tip is transparent to light of the wavelength $150\; \textrm{nm} < \lambda < 4000 \; \textrm{nm}$, thus not cutting into the UV-spectrum of the deuterium lamp. The work function of 
gold is $4.83\pm 0.02$~eV \cite{And59}. Therefore, only electrons with excess energies of up to $E_{\textrm{max}} = 6.7 \;\textrm{eV} - 4.83 \;\textrm{eV} = 1.87 \;\textrm{eV}$ can be released. The gold plated tip is supplied with a voltage of $U_{\textrm{tip}}=-18$~kV. The photoelectrons are finally accelerated to the energy $qU_{\textrm{tip}}=18$~keV in forward direction by a blind (a) on ground potential.

The electron source is mounted on a manipulator which allowed us to move it on a sphere up to $\alpha \pm 19^{\circ}$ into horizontal and vertical direction (cf. Fig. \ref{fig::prespec_setup}). This corresponds to a motion for the gold plated tip on a radius of 1.06 m around the point with the pre-spectrometer coordinates $z=-2.4$~m and $r=0$~m, 0.25 m behind the center of the source magnet (cf. Fig. \ref{fig::prespec_setup}). The intensity stability of the electron source was measured to better than 0.2 \% per hour. Fig. \ref{fig::E-GUN} shows the cross-section of the E-GUN. The reader should keep in mind that the electric field near the gold plated tip (b) is strong
and there is a ground blind (a) mounted in front of the tip. Together with the electric and magnetic fields inside the E-GUN and the detector, these two components play an important role in our 
data analysis (Section \ref{sec::simulation}).\par

The pre-spectrometer detector is a quadratically segmented
silicon PIN diode with 64 pixels of equal size and properties. It has an overall sensitive area of $16\;\textrm{cm}^2$ \cite{Wue06}. It is a predecessor of the final KATRIN detector and was manufactured with the same processing techniques. For simulations of the detector response, the energy resolution $\Delta E_{\textrm{FWHM}}$ and the dead layer thickness $\lambda$ have to be known. The detector system exhibited a measured average $\Delta E_{\textrm{FWHM}}\approx 3.5 \;\textrm{keV}$. The thickness of the dead layer was determined as $\lambda=119 \pm 3\;\textrm{nm}$ and $\lambda=109 \pm 3\;\textrm{nm}$ by using two independent experimental 
techniques \cite{Wal06}. The detector is located at $z=2.3$~m (15 cm behind the center plane of the detector magnet) at $B=3.4\;\textrm{T}$ and at ground potential (Fig. \ref{fig::prespec_setup}). 
If the detector is centered on the PS axis, its area corresponds to 28.5 \% of KATRIN's magnetic flux tube. In our measurements, the detector was adjusted laterally so that  only a single pixel was hit and the data analysis was made with this single pixel. The energy calibration 
of the pixel was used to select events in the region of interest from 15 keV to 21 keV.\par

For our measurements, the inner electrodes (E) and (F) (cf. Fig. \ref{fig::prespec_setup}) were put on a voltage of $U_{\textrm{electrode}} -  U_{\textrm{tank}} =  -500 \pm 0.1\;\textrm{V}$ with respect to the 
PS tank using a voltage supply (Canberra 3101/2) mounted inside a rack on tank potential. The tank voltage itself ($U_{\textrm{tank}}$ between -0.5 kV and -17.5 kV) was supplied 
by another voltage source (FUG HCN 140M-35000). The gold plated tip of the E-GUN was supplied with a constant voltage of
$U_{\textrm{tip}}=-18$~kV by a high voltage supply (FUG HCN 35-35000). The accuracy of the voltage difference determination between the tank and the gold plated tip was better than 10 V.  

\section{Measurements}
\label{sec::measurements}

We performed six measurement series with $E=18$~keV electrons (Tab. \ref{tab::summary_of_measurements}). The
PS  solenoids were set to the KATRIN design value of $B_{\textrm{sol}}=4.5\;\textrm{T}$ (Figs. \ref{fig::katrin_overview} and \ref{fig::prespec_setup}) and half this value $B_{\textrm{sol}}=2.3\;\textrm{T}$. As the cyclotron length grows with $1/B$ (Eq. \ref{eq::cyclotron_length}), deviations from the ideal transmission properties described in Section \ref{sec::mac_e_principle} are 
more probable for the decreased B-field $B_{\textrm{sol}}=2.3\;\textrm{T}$. The electrons pass the PS on a radius $r(z)$, which encloses a constant magnetic flux $\phi_{\textrm{enc}}$ (in a homogeneous B-field one has 
$\phi_{\textrm{enc}} \approx \pi \cdot r^2(z) \cdot B(z)$). For each B-field, the PS transmission was measured for three different angular positions ($\alpha \in \{ 0^{\circ}, 15^{\circ}, 19^{\circ}\}$) of the E-GUN (cf. Fig. \ref{fig::prespec_setup}).
With the given angles $\alpha$, electrons pass the central plane of the PS at the radii $r_{\textrm{cen}} \in \{ 0\; \textrm{cm}, 41.4\; \textrm{cm}, 52.3\; \textrm{cm} \}$. At the KATRIN design 
value $B_{\textrm{sol}}=4.5$~T, the corresponding fraction $f= \phi_{\textrm{enc}} / \phi_{\textrm{KAT}}$ of the enclosed KATRIN flux tube $\phi_{\textrm{KAT}} = 191 \;\textrm{T}\cdot \textrm{cm}^2$ is $f  \in \{ 0\; \%, 45\; \%, 72\; \% \}$.

\begin{table}[!h]
\begin{center}
\caption{Summary of measurements.}
\label{tab::summary_of_measurements}
\begin{tabular}{cllll}
\hline
$B_{\textrm{sol} }\;\textrm{(T)}$	&	 $r_{\textrm{cen}}$ (cm)					&	scans 	&	measurement time (s)	&	events at $qU_{\textrm{electrode}}=17.5\;\textrm{keV}$       \\
\hline\hline
					& 	0					&	5		&	478					&	$1.9 \cdot 10^{6}$		\\
	4.5				& 	41.4					&	3		&	287					&	$1.0\cdot 	10^{6}$		\\
					&	52.3					&	4		&	380					&	$1.3 \cdot 10^{6}$		\\
\hline
					&	0					&	3		&	287					& 	$1.0\cdot 	10^{6}$		\\
	2.3				&	41.4					&	4		&	414					&	$1.0\cdot 	10^{6}$		\\
					&	52.3					&	2		&	192					&	$0.4\cdot 	10^{6}$		\\
\hline
\end{tabular}
\end{center}
\end{table}

\begin{figure}[!htbp]
\begin{center}
\includegraphics[width=0.95\textwidth]{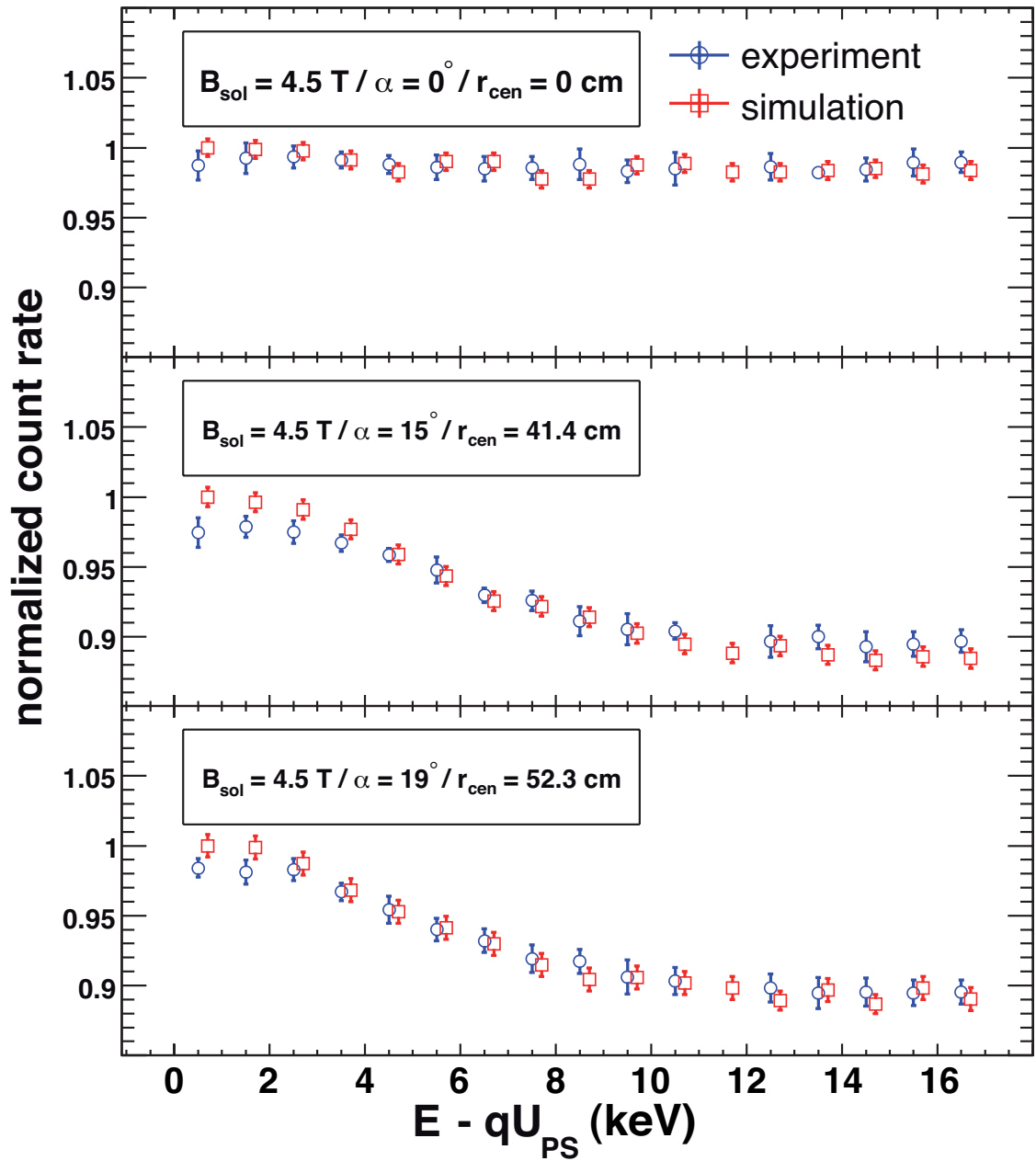}
\caption{Measurement (blue circles) and simulation (red squares) for $B_{\textrm{sol}}=4.5\;\textrm{T}$. The simulation was normalized so that the rate at $E_{\textrm{sur}} = E - qU_{\textrm{PS}}=0.5$~keV is 1.
The experimental rate was normalized so that its average rate is equal to the average of the simulated rate. Only simulation values where a measurement exists were considered for this average.
The simulation points are shifted to the right by 0.2~keV to make the points distinguishable.
The plot shows that our simulation and our measurements are compatible at the percent level.}
\label{fig::comparison_fullfield}
\end{center}
\end{figure}

\begin{figure}[!htbp]
\begin{center}
\includegraphics[width=0.95\textwidth]{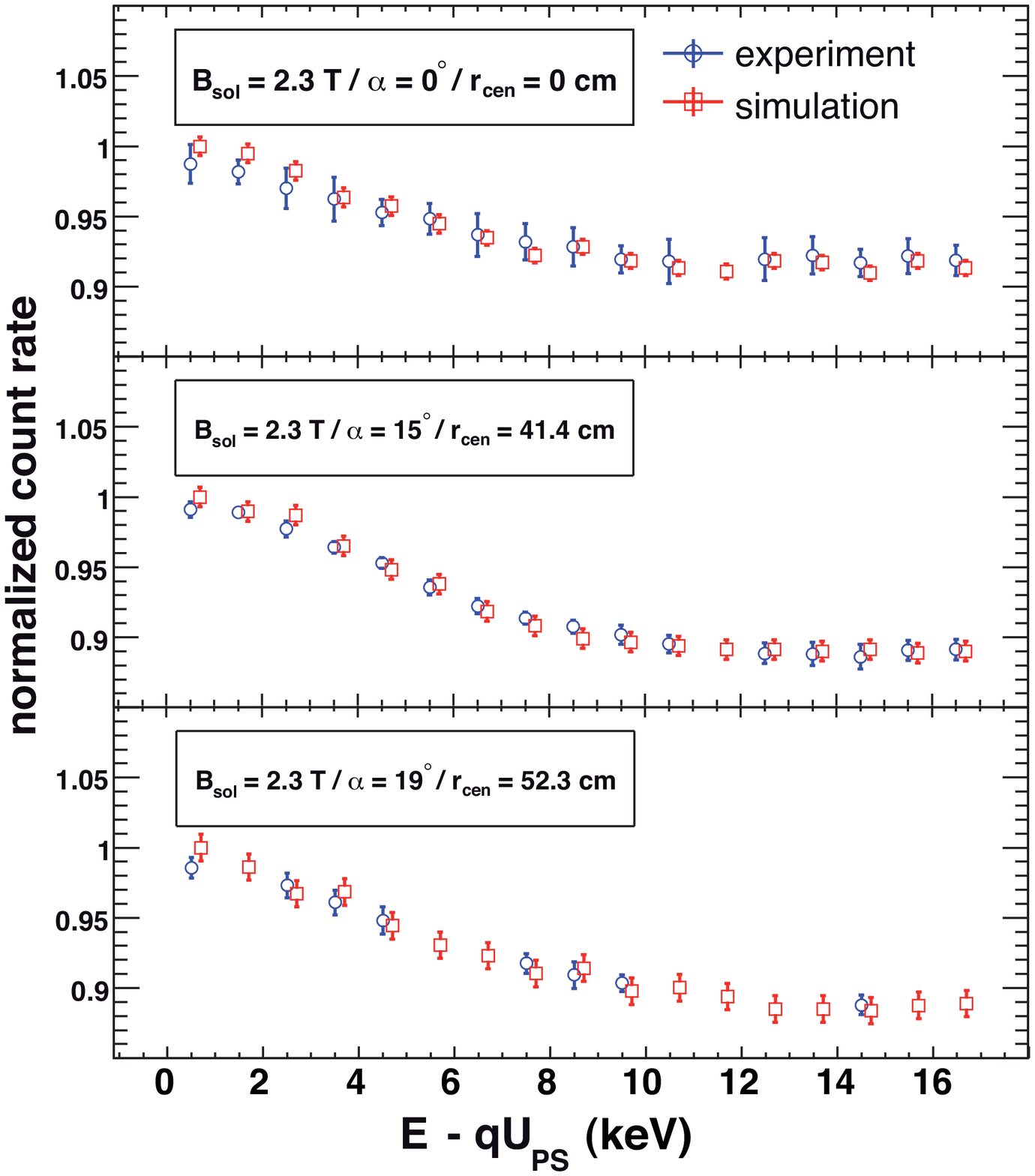}
\caption{Measurement (blue circles) and simulation (red squares) for $B_{\textrm{sol}}=2.3\;\textrm T$. The simulation was normalized so that the rate at $E_{\textrm{sur}} = E - qU_{\textrm{PS}}=0.5$~keV is 1.
The experimental rate was normalized so that its average rate is equal to the average of the simulated rate. Only simulation values where a measurement exists were considered for this average.
The simulation points are shifted to the right by 0.2~keV to make the points distinguishable. At $\alpha=19^{\circ} \; / \;r_{\textrm{cen}}=52.3\;\textrm{cm} $ we measured at fewer filter potentials.
The plot shows that our simulation and our measurements are compatible at the percent level.}
\label{fig::comparison_halffield}
\end{center}
\end{figure}

In each measurement series, the PS filter energy $qU_{\textrm{PS}}$ was stepped repeatedly from 
1 keV towards 18 keV and back to 1 keV, using identical time intervals. Combining the counts from the ramp up and the ramp down eliminates a possible linear drift in 
the emission rate of the E-GUN. Each detector run at a constant potential lasted for about $R=48\;\textrm{s}$. The procedure was repeated up to five 
times (see 'scans' in  Tab. \ref{tab::summary_of_measurements}), resulting in an overall measurement time of e.g. $478\;\textrm{s}\approx \textrm{scans}\cdot 2\cdot R$ for
the measurement with $B_{\textrm{sol}}=4.5\;\textrm{T}$ and $r_{\textrm{cen}}=0\;\textrm{cm}$. From the runs with the identical voltages, $r_{\textrm{cen}}$ and B-field settings, electron events were summed and divided by the overall measurement time to obtain an average electron rate. The detector was adjusted laterally before the start of the measurements so that only a single detector pixel was hit. Only events within the region of interest  $15\;\textrm{keV} < E < 21\;\textrm{keV}$ and from this single detector pixel were counted. The comparison of our six measurement series at keV surplus energies with our simulations are shown in Figs. \ref{fig::comparison_fullfield} and \ref{fig::comparison_halffield}. Except for a single measurement ($B_{\textrm{sol}}$\;=\; 4.5~T, $r_{\textrm{cen}}=0\;\textrm{cm}$), the detector rates at positive surplus energies always decrease with growing surplus energy $E-qU_{\textrm{PS}}$. The explanation of this observation is given in the section \ref{sec::simulation}.

\section{Simulation Tools}
\label{sec::simulation_tools}

The analysis of our experimental data is done by comparing it to computer simulations.
The main components of our simulations are:
(i), electric and magnetic field computations;  (ii), electron tracking in vacuum;
(iii), electron scattering with ${\rm H}_2$ molecules; (iv), electron tracking in silicon.
For the electric field calculations (axisymmetric and three-dimensional, with wires) we used the
boundary element method \cite{Corona,Glu11a}. In order to speed up the simulation in axisymmetric regions, we
employed the zonal harmonic expansion method \cite{Glu11b}. This turned out to be useful also 
for the E-GUN -- pre-spectrometer geometry, which is not axially symmetric but consists of two 
separated, locally axisymmetric 
regions (E-GUN region and pre-spectrometer region).
The  zonal harmonic expansion method was also used for magnetic field computations
\cite{Glu11c}.
For the electron tracking in vacuum, 
the exact relativistic equation of motion of the electron with  Lorentz force 
was employed \cite{Glu11a}, using 
an explicit 8th order Runge-Kutta method to solve the ordinary differential equations \cite{Verner}.
The electron-${\rm H}_2$ scattering code contains total and differential cross sections 
\cite{For73,Hwa96,Kaw94,Liu73,Liu87,Liu85,Rud93,Taw90,Traj83} and Monte Carlo generation 
algorithms for elastic, electronic excitation and ionization collisions of
electrons with ${\rm H}_2$ molecules  \cite{Glu11a}.
Electron detection,
electron energies deposited in the sensitive volume of the silicon detector, the detector dead layer, and
electron backscattering at the detector are modeled by a Monte Carlo C++ code
(KESS: KATRIN Electron Scattering in Silicon),
which is based 
on detailed studies \cite{Chaa,Chab,Chac,Cha10} and agrees well with experimental data \cite{Ren10}.

Our original field calculation, 8th order Runge-Kutta tracking and e-${\rm H}_2$ scattering C codes \cite{Glu11b,Glu11c,Glu11a} have been rewritten into C++ and
integrated into the global KATRIN C++ simulation framework
 'Kassiopeia' \cite{Kassio}. 
KESS \cite{Ren11} has also been integrated into Kassiopeia.
We have used both the original C codes and the new Kassiopeia C++ code for the simulations
of our paper.

\section{Simulation and Analysis}
\label{sec::simulation}

\begin{figure}[!htbp]
\begin{center}
\includegraphics[width=\textwidth]{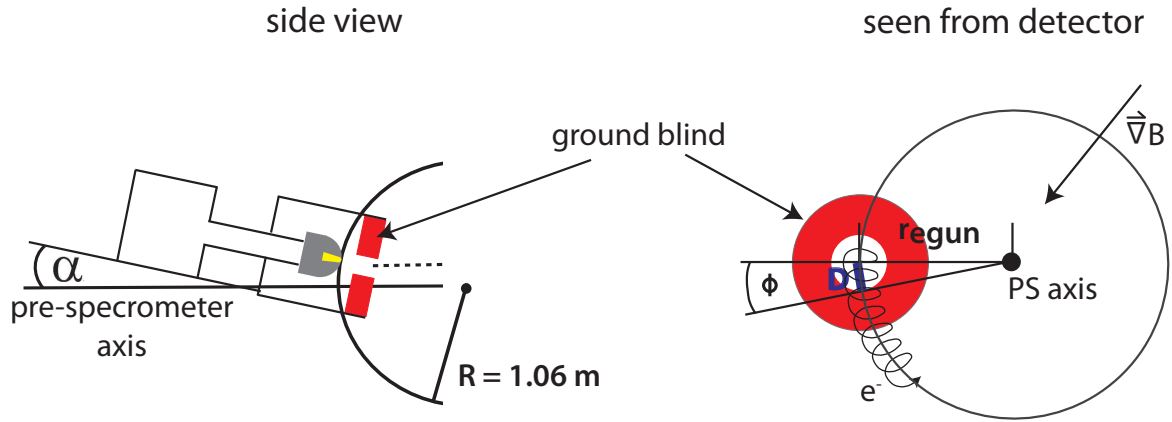}
\caption{The azimuthal drift in the pre-spectrometer. Left: Geometry of the E-GUN. Right: The azimuthal drift seen from the detector side.
The electron starts in the centre of the bore in the ground blind. Towards the axis of the pre-spectrometer, the B-field is stronger and the 
curvature of the electron trajectory is larger, resulting in an overall $\vec{B} \times \vec{\nabla} B$ drift around the pre-spectrometer axis (\ref{eq::drift_perp}). If the drift is large enough, the
backscattered electron hits the ground blind on its way back from the detector.}
\label{fig::reflection_principle}
\end{center}
\end{figure}

Ignoring the influence of the E-GUN and the detector, the PS transmission at keV surplus energies in our codes \cite{Glu11b,Glu11c,Glu11a} is always 100\%: In order to show this, we started 130 electrons with uniformly distributed polar- and azimuthal angles for $B_{\textrm{sol}} \in \{ 2.3\;\textrm{T}, 4.5\;\textrm{T} \}$, $U_{\textrm{electrode}}=U_{\textrm{tank}}- 0.5\;\textrm{kV}\in \{-1.5\;\textrm{kV},-2.5\;\textrm{kV},..,-17.5\;\textrm{kV} \}$ and $r_{\textrm{start}}\in \{ 0\;\textrm{mm}, 26\;\textrm{mm}, 37\;\textrm{mm} \}$ in the entry side magnet ($z=-2.15$~m) of the PS. For every simulated electron, the 
exit condition $(z \geq 2.15\;\textrm{m} \; \textrm{and} \; r\leq 37\;\textrm{mm})$ was reached.
In the following we show that the measured decrease of detector rate at positive surplus energy $E - qU_{\textrm{PS}}$ is not caused by a loss of transmission in the PS, but by the special setup (Fig. \ref{fig::prespec_setup}) used for the measurements. The transmission probability of 100\% can only be confirmed indirectly with an uncertainty at the percent level via the comparison with a simulation. 
The essential ingredients to explain our data are:

\begin{enumerate}
	\item A constant PS transmission probability of 100\%
	\item Electron backscattering at the detector
	\item $\vec{B}\times \vec{\nabla}B$-drift of electrons proportional to $E_{\textrm{kin}}$ (see (\ref{eq::drift_perp}))
	\item Reflection of electrons in the electric field of the E-GUN
	\item The loss of backscattered electrons hitting the ground blind in front of the E-GUN (Figs. \ref{fig::E-GUN} and \ref{fig::reflection_principle})
\end{enumerate}

Electrons impinging on the detector with an incident energy $E_{\textrm{inc}}=18$\,keV and an incident angle $\theta_{\textrm{inc}}=0^\circ$ have a probability of about 20\% to be backscattered, and higher incident angles further increase this probability \cite{Dar75,Dre70}. Most of the backscattered electrons have lost energy in the detector and are again reflected by the filter potential $qU_{\textrm{PS}}$ in the PS or by the magnetic mirror effect towards the detector. Backscattered electrons having deposited less than $E - qU_{\textrm{PS}}$ in the detector retain enough energy to pass the filter potential $qU_{\textrm{PS}}$ of the PS in backward (towards E-GUN) direction. They can enter the E-GUN through the opening in the ground blind (cf. Fig. \ref{fig::E-GUN}) and can be electrically reflected towards the detector again. This process continues until all energy is finally deposited inside the detector. Even at high surplus energy $ E - qU_{\textrm{PS}}$, the count rate at the detector is constant. Flight times in the PS are of the order of 10 ns which is far below the $\mu$s shaping time of the DAQ. Therefore, only electrons with large energy losses in the deadlayer will deposit an energy lower than the region of interest in the sensitive volume. Since all electrons hitting the detector have the same energy and same angular distribution for all settings of $U_{\textrm{electrode}} = U_{\textrm{tank}} - 0.5\;\textrm{kV}$, the count rate does not depend on this voltage setting. This explains the measurement at $B_{\textrm{sol}} = 4.5\,\textrm{T}$ and $r_{\textrm{cen}}=0\;\textrm{cm}$ ($\alpha=0^{\circ}$), where no loss in count rate is observed.\\

The description above is also valid for the off-axis E-GUN settings with $r_{\textrm{cen}} \in \{ 41.4 \;\textrm{cm}, 52.3 \;\textrm{cm} \} $ ($\alpha \in \{ 15^{\circ}, 19^{\circ} \}$). 

The axial rotation (\ref{eq::drift_perp}) can cause the electron to eventually hit the ground blind (Fig. \ref{fig::reflection_principle}), depending on $\vec{B}$, the electron energies $E_{\perp}$ and $E_{\parallel}$ and the total path length. The axial rotation is only dependent on the energy and has the same sense for a forward (towards detector) and backward (towards E-GUN) pass of the PS and is therefore adding up at each pass of the PS.
The path for the electrons in the pre-spectrometer is elongated by reflections at the detector, at the field of the E-GUN, at magnetic fields and at the spectrometer potential and can thus be multiples of the spectrometer length.
The higher the surplus energy, the higher the probability for a backscattered electron to overcome the spectrometer potential after energy deposits in the detector. 
The higher the surplus energy, the larger the axial rotation which guides electrons onto the ground blind. Therefore, the count rate in the region of interest decreases with higher surplus 
energy $E_{\textrm{sur}} = E_{\textrm{0}}-qU_{\textrm{PS}}   $. With this, all measurements at $B_{\textrm{sol}}=4.5$\,T can be explained.\\

For $B_{\textrm{sol}}=2.3$\,T a loss in count rate is observed for all E-GUN settings, including the one at  $r_{\textrm{cen}}=0$~cm ($\alpha=0^{\circ}$). It is not possible to explain this effect with backscattering and the $\vec{B} \times \vec{\nabla}  |\vec{B}|$ drift alone. 
The electric field gradient in the E-GUN is large compared to the PS, since the potential difference of 18\,kV is applied across a distance of only a few cm. Together with the 50\% lower magnetic field in the center of the magnets, this can lead to a non-adiabatic transport in the E-GUN region. At $B_{\textrm{sol}}=2.3\;\textrm{T}$, the E-GUN is located at a magnetic field of $B\approx0.02$\,T. Thus, a backscattered electron entering the E-GUN through the ground blind has a probability to change its angle towards the magnetic field line non-adiabatically. Depending on the new angle and the electron energy, it can be trapped between the E-GUN and the closest magnet or the spectrometer potential. Thus, a loss in count rate in the region of interest will also be observed for measurements with $r_{\textrm{cen}}$ ($\alpha=0^{\circ}$). This assumption is therefore able to explain the measurement for $B_{\textrm{sol}}=2.3$\,T and $r_{\textrm{cen}}$ ($\alpha=0^{\circ}$).

In our simulations, the electrons were started with a uniform random kinetic energy of $0<E<2$\,eV, and were uniformly distributed on a disc with diameter $d=1$\,mm in front of the actual gold tip. An 
angular distribution, $\theta=\arcsin(R_1)$ from \cite{Pei02} and $\varphi=2\pi R_2$ with uniformly distributed random numbers $R_1$, $R_2$ $\in [0,1[$ was used. The time an electron travels between two subsequent detector hits is more than two magnitudes smaller than the DAQ shaping time. This means, subsequent hits are analyzed by the DAQ as one hit. Therefore, each energy deposition in the sensitive detector volume per electron was summed up, even for electrons with multiple detector entries. As in the experimental data analysis, all electrons with energies $15<E<21$\,keV were counted.\par

\newpage

Possible exit conditions of the simulation were: 

\begin{itemize}
\item Electron hits E-GUN blind (cf. Fig. \ref{fig::E-GUN} and Fig. \ref{fig::reflection_principle}).
\item Electron energy lower than 100\,eV. It is no longer able to pass the dead layer of the detector. 
\item Electron was reflected more than 20 times in the PS (trapping).
\end{itemize}

Figures \ref{fig::comparison_fullfield}  and \ref{fig::comparison_halffield} show the normalized simulation and experimental results. 
They show that our simulation can explain the experimental rates at the percent level. The agreement between experiment and simulation implies that the PS has 100\% transmission probability 
for electrons with keV surplus energy as long as $B_{\textrm{sol}}$ is not smaller than 2.3~T (half the KATRIN design value \cite{Ang04}). 

% Ferenc's Kapitel nach Bearbeitung von Herrn Otten's Kommentaren

\section{Operating the PS at reduced zero filter energy} 
\label{sec::pre_spec_modes}

The previous section proves experimentally that the PS can be operated at reduced or even zero filter energy without essential transmission losses, as long as $B_{\textrm{sol}}$ is not smaller than 2.3 T.  In this section we discuss more generally, with the help of simulations, the operation of PS with zero potential. We show first that, with $U_{\textrm{PS}}=0$ ,
  the background due to positive ions created by beta electrons is expected to be far below the KATRIN requirement of $b=10$ mHz, and second that the electron motion through the PS and MS is fully adiabatic.

Lowering the filter energy $qU_{\textrm{PS}}$ of the PS, the depth of the natural Penning trap, created by the PS and MS retarding potentials $U_{\textrm{PS}}$ and $U_{\textrm{MS}}$ and the magnetic field $B_{\textrm{sol}}$, between the PS and MS decreases, and the corresponding background level is expected to decrease, too. On the other hand, there is another background component that increases with decreasing $qU_{\textrm{PS}}$: more $\beta$-electrons reach the entrance of the main spectrometer.
These electrons are not able to produce direct background, but they create positive ions through ionizing collisions with
  the residual gas molecules. These ions can fly deep into the main spectrometer, and they can produce low energy electrons there, either in the residual gas or at the inner surface of the main spectrometer electrodes.
Some of these electrons can hit the detector, and so we obtain background events.
We will present below a quantitative estimate for the maximal value of this background component, in case of zero PS potential.

Inside the WGTS, $10^{11}$ electrons are produced through tritium beta decays each second. About 20 \% of them move through the transport system and reach the  entrance of the main spectrometer in the case of $qU_{\textrm{PS}}=0\;\textrm{keV}$. Along their way, many of these electrons have ionizing collisions with residual gas molecules and thus produce positive ions. In the absence of any hindrances, all these ions enter the main spectrometer.
In order to reduce the background rate of these ions, we plan to use an ion-blocking electrode near the center of the main spectrometer entrance magnet. Since the positive ions created by ionizing collisions have small kinetic energy (below few eV), an accordingly small potential barrier created by an ion-blocking electrode would already
  prevent these ions from entering the MS and thus from producing background.

Nevertheless, the ions
created in the region between the ion-blocking electrode and the high-voltage area of the MS are able to enter the MS. We have computed the ion creation rate in this region by detailed trajectory simulations. First, we generated electrons at the center of the MS entrance magnet by using the tritium $\beta$-decay Fermi spectrum and an isotropic angular distribution.
These electrons were tracked until their reflection at the MS filter potential and then back to their starting point. The ionization probability of an electron was computed by summing the differential ionization probabilities $d{\cal P}_{\rm ion}=\sigma_{\rm ion}(E_{\rm kin}) \cdot n \cdot dl$, where $\sigma_{\rm ion}(E_{\rm kin})$ denotes the ionization cross section of electrons with the residual gas molecules, as function of the electron kinetic energy $E_{\rm kin}$, $n$ is the number density of the residual gas, and $dl$ the differential pathlength.
In our simulations, we assumed molecular hydrogen for the residual gas, and we used the ${\rm e}-{\rm H}_2$ ionization cross section formulas of \cite{Hwa96,Liu73} (they are in good agreement with measured ${\rm e}-{\rm H}_2$ cross section values).
Assuming $p=10^{-11}\; {\rm mbar}$ pressure and room temperature, the number density is
$n=2.4 \cdot 10^{11}$ ${\rm m}^{-3}$. According to our calculations (simulation of 1000 electron tracks), the average ionization cross section 
is $\sigma_{\rm ion}=10^{-21}$ ${\rm m}^{2}$, the average electron pathlength is $l=1.5$ m, and the average ionization probability of an 
electron is $ {\cal P}_{\rm ion}=3.6\cdot 10^{-10}$.
Using the $2\cdot 10^{10}\,{\rm s}^{-1}$ $\beta$-electron intensity, we obtain a positive ion creation rate of roughly
 $\dot{N}_+\approx 7$ ${\rm s}^{-1}$ in the region between the ion-blocking electrode and the high potential domain of the MS.

If an electron is scattered towards large polar angles $\theta$ (remember $E_{\perp}=E_{\textrm{kin}} \cdot \sin^2 \theta$), it can become
 trapped in a hybrid trap near the entrance of the MS:
if this electron moves towards the entrance magnet of the MS coming from inside the MS, $\theta$ increases adiabatically 
until $\theta=90^{\circ}$ is reached, and the electron starts to move towards the MS again. The MS magnet thus establishes a magnetic
 mirror for these electrons. Inside the MS, the electron is electrically reflected by the MS filter potential generated by the MS electrode 
system \cite{Pra12}.
In order to compute the ion creation rate due to these trapped electrons, we simulated $10^8$ electrons. We used our custom C 
codes \cite{Glu11a,Glu11b,Glu11c} to compute the electromagnetic fields, the trajectories and the ${\rm e}-{\rm H}_2$ scattering.
  The result of these simulations is the following: the average trapping probability of the beta electrons in the hybrid trap
is $3\cdot 10^{-11}$, and the number of ions created by a trapped electron is smaller than 5
(the trapped electrons can leave
 the trap by scattering and by energy loss due to synchrotron radiation). From these numbers and from the above beta intensity
we get an ionization rate that is smaller than 3 ${\rm s}^{-1}$ .
Therefore, the ion creation rate due to these trapped electrons is
smaller than due to the free (non-trapped) electrons.

The  ions  with roughly  $\dot{N}_+\approx 10$ ${\rm s}^{-1}$ rate enter the MS,
they will be accelerated to about 18.57 keV kinetic energy, and due to the small (few Gauss) magnetic field, their motion inside the MS is completely non-adiabatic: they move on a straight line, until they hit the spectrometer tank.
During this motion, they can suffer ionizing collisions with the residual gas. The ionizing collision cross section of ${\rm H}^+$ and ${\rm H}_2^+$ ions of 18 keV kinetic energy with ${\rm H}_2$ molecules is about $\sigma_{\rm ion}^+=2\cdot 10^{-20}$ ${\rm m}^2$ \cite{Tab00}. 
Assuming $l_+=20$ m pathlength for the positive ions inside the main spectrometer tank, the secondary electron creation rate due to ionizing collisions of the positive ions with ${\rm H}_2$ molecules is $\dot{N}_e=\dot{N}_+ \sigma_{\rm ion}^+\,  l_+\, n =10^{-6}$ ${\rm s}^{-1}$, corresponding to 0.001 mHz background level.
This background increases quadratically with the residual gas pressure, so with $p=10^{-10}\; {\rm mbar}$ the background rate would be 0.1 mHz.

Another background possibility is the following: the ions hit the MS tank with high velocity, and these impact events are connected with secondary electron emission from the surface. One ion can eject more than 1 electron; let us assume that this multiplication number is 10. Then we get a secondary electron emission rate of 100 ${\rm s}^{-1}$ from the tank surface. This is several orders of magnitude smaller than we expect from cosmic ray muons and environmental radioactivity.
Thanks to the magnetic shielding  of the approximately axisymmetric magnetic field and to the electric shielding of the wire electrode, only a very small proportion of these electrons is expected to reach the detector; extrapolating the experimental data of the Mainz neutrino mass spectrometer, this proportion could be about $10^{-7}$. With this suppression factor, we get 0.01 mHz background level from these electrons.

For both scenarios we obtain a background level caused by positive ions which is several orders of magnitude smaller than the $b=10$~mHz background value that would be acceptable for the KATRIN experiment \cite{Ang04}. 
Therefore, our simulations show that the PS could be used with zero potential, without any significant background increase due to the positive ions produced by the beta electrons.

With zero or small PS filter energy, the signal electrons
($E_{\textrm{0}}=18.57$~keV) have a high surplus energy \mbox{$E_{\textrm{sur}} =  E_{\textrm{0}} - qU_{\textrm{PS}}$} inside the PS. Due to the relatively small magnetic field in the middle of the PS ($B_{\textrm{ana,PS}}=0.016$~T), it could happen, in principle, that the motion of these electrons is not adiabatic. We have checked the adiabaticity behaviour of the electrons with detailed trajectory simulations. For this purpose, we started the electrons in the KATRIN source (WGTS) at various points with various polar and azimuthal direction angles, and we tracked them as far as the main spectrometer analyzing plane. We defined the starting kinetic energy of the electrons with the following procedure: first, using the starting point, direction vector and a first estimate for the
  transmission energy, we computed the guiding center point corresponding to the starting point. Then, we simulated the magnetic field line from the guiding center point until the MS analyzing plane. 
Using the electric potential and magnetic field values at the two endpoints of this field line, and the starting polar angle, we calculated the adiabatic transmission energy: in the adiabatic approximation, the electron has zero longitudinal energy $E_{\parallel}$ in the analyzing plane (cf. Fig. 
\ref{fig::mac_e_principle}) if it starts with this energy (electrons starting below or above this energy are reflected or transmitted, respectively). We defined the starting kinetic energy of our electrons as the above adiabatic transmission energy plus a small surplus energy ($E_{\rm sur}=E - qU_{\textrm{PS}}$= 3 meV).
  If the electron motion is fully
adiabatic, $E_{\parallel}$ has to be precisely equal to $E_{\rm sur}$ in the analyzing plane. The main result of our simulations is the following: both for a PS filter energy of $qU_{\textrm{PS}}=0$~keV and $qU_{\textrm{PS}}=18.3$~keV,
and for all starting parameters, using the standard KATRIN magnetic field values (3.6 T in WGTS, etc.), $|E_{\parallel}-E_{\rm sur}|$
(computed by exact tracking) is in the MS analyzing plane on the average 0.2 meV, which is four orders of magnitude smaller than the resolution $\Delta E_{\textrm{MS}}=0.93$~eV of the KATRIN MS.
Therefore, we can say that the motion of signal electrons in the KATRIN system is practically adiabatic, even with
$U_{\textrm{PS}}=0\;\textrm{kV}$; deviations from adiabaticity have a negligible effect to the KATRIN transmission function. Note that that the electron motion is approximately adiabatic even if the magnetic field in the whole KATRIN system is half of its standard
design value; in this case, $|E_{\parallel}-E_{\rm sur}|$ is in the MS analyzing plane on the average 0.6 meV.

%The main result of our simulations is the following: both for a PS filter energy of $qU_{\textrm{PS}}=0$~keV and $qU_{\textrm{PS}}=18.3$~keV, and for all starting parameters, using the standard KATRIN magnetic field values (3.6 T in WGTS, etc.), $|E_{\parallel}-E_{\rm sur}|$ (computed by exact tracking) is smaller than 0.1 meV in the MS analyzing plane, which is four orders of magnitude smaller than the resolution $\Delta E_{\textrm{MS}}=0.93$~eV of the KATRIN MS.
%Therefore, we can say that the motion of signal electrons in the KATRIN system is practically adiabatic, even with $U_{\textrm{PS}}=0\;\textrm{kV}$; deviations from adiabaticity have a negligible effect to the KATRIN transmission function.
%Note that that the electron motion is approximately adiabatic even if the magnetic field in the whole KATRIN system is half of its standard design value; in this case, with the above electron parameters,
%  $|E_{\parallel}-E_{\rm sur}|$ is smaller than 3 meV
% in the MS analyzing plane.

\begin{figure}[!ht]
\begin{minipage}[t]{8cm}
\centerline{\includegraphics[height=3.3cm]{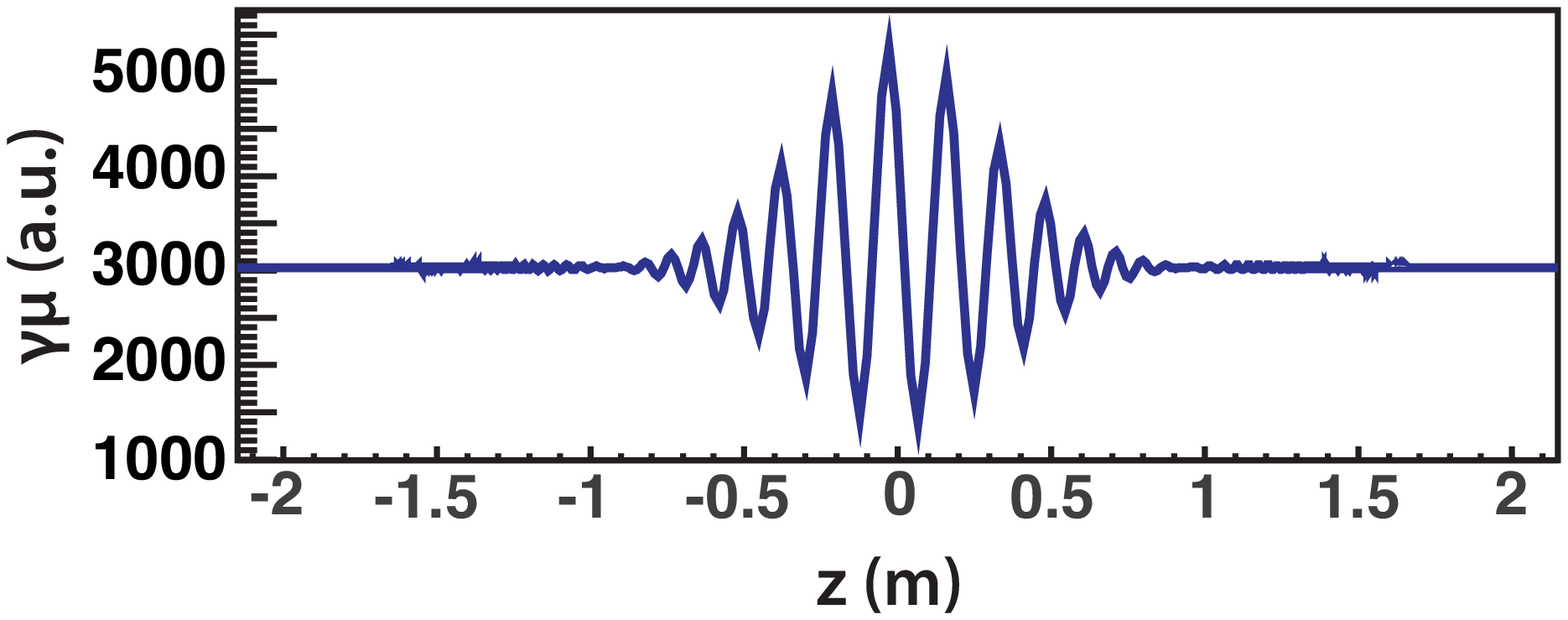}}
\caption{Oscillations of $\boldmath{\gamma\mu}$ and regain of original
value: $B_{\textrm{sol}}=4.5$~T, $U_{\textrm{PS}}=-1.5$~kV, $E=18$~keV, $z_{\textrm{start}}=-2.15\;\textrm{m}$,
$\varphi_{\textrm{start}}=60^{\circ}$ (azimuthal angle), $\theta_{\textrm{start}}=72^{\circ}$ (polar angle) and $r_{\textrm{start}}=26\;\textrm{mm}$. The PS solenoids are at $z=-2.15\;\textrm{m}$ and $z=+2.15\;\textrm{m}$. The radius of the KATRIN flux tube in solenoids is 37~mm at $B_{\textrm{sol}}=4.5$~T.} \label{fig::adi_regeneration} \end{minipage} \hfill \begin{minipage}[t]{8cm} \setlength{\fboxsep}{0mm} \centerline{\includegraphics[height=3.3cm]{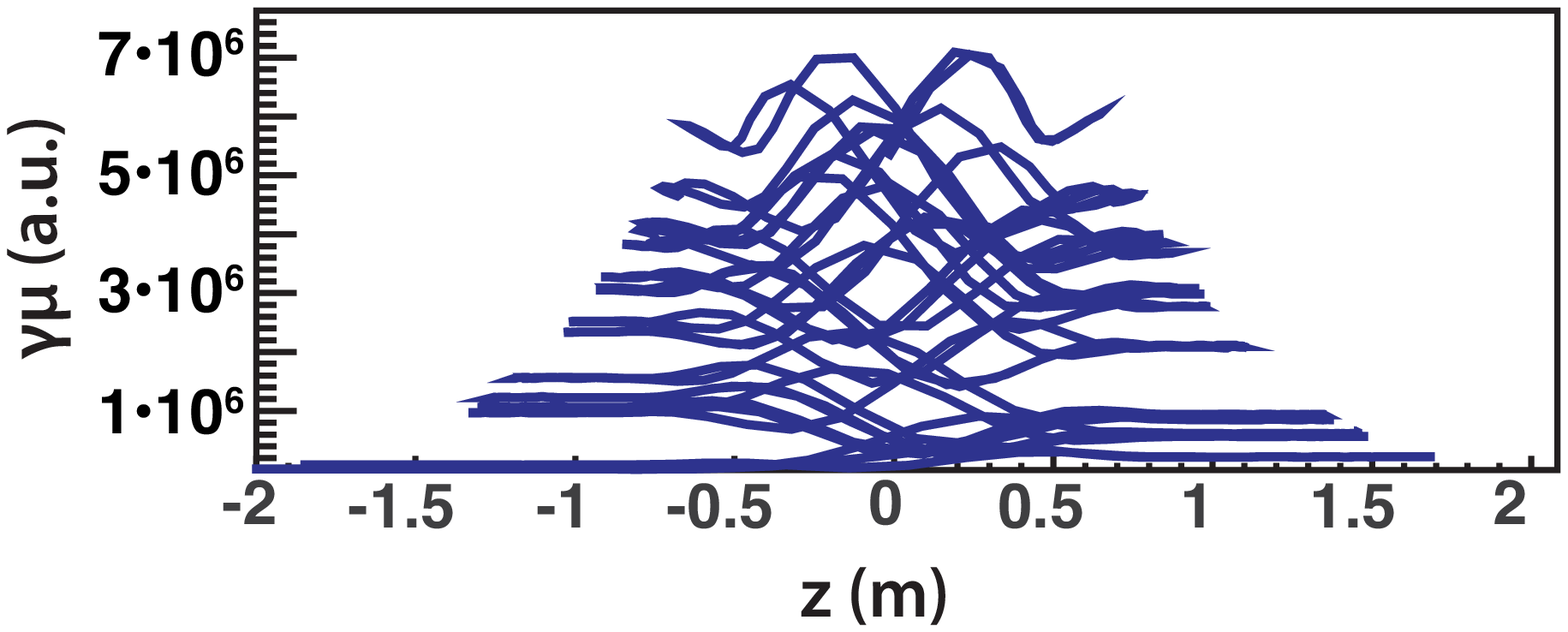}}
\caption{Evolution of  $\boldmath{\gamma\mu}$ for a trapped electron: 
$B_{\textrm{sol}}=0.57$~T, $U_{\textrm{PS}}=-1.5$~kV, $E=18$~keV, $z_{\textrm{start}}=-2.15\;\textrm{m}$,
$\varphi_{\textrm{start}}=30^{\circ}$ (azimuthal angle), $\theta_{\textrm{start}}=72^{\circ}$ (polar angle) and $r_{\textrm{start}}=26\;\textrm{mm}$. The PS solenoids are at $z=-2.15\;\textrm{m}$ and $z=+2.15\;\textrm{m}$. The radius of the KATRIN flux tube in solenoids is 37~mm at $B_{\textrm{sol}}=4.5$~T.} \label{fig::adi_change_trapped} \end{minipage} \end{figure}

\begin{figure}[!htbp]
\begin{minipage}[t]{8cm}
\centerline{\includegraphics[height=3.3cm]{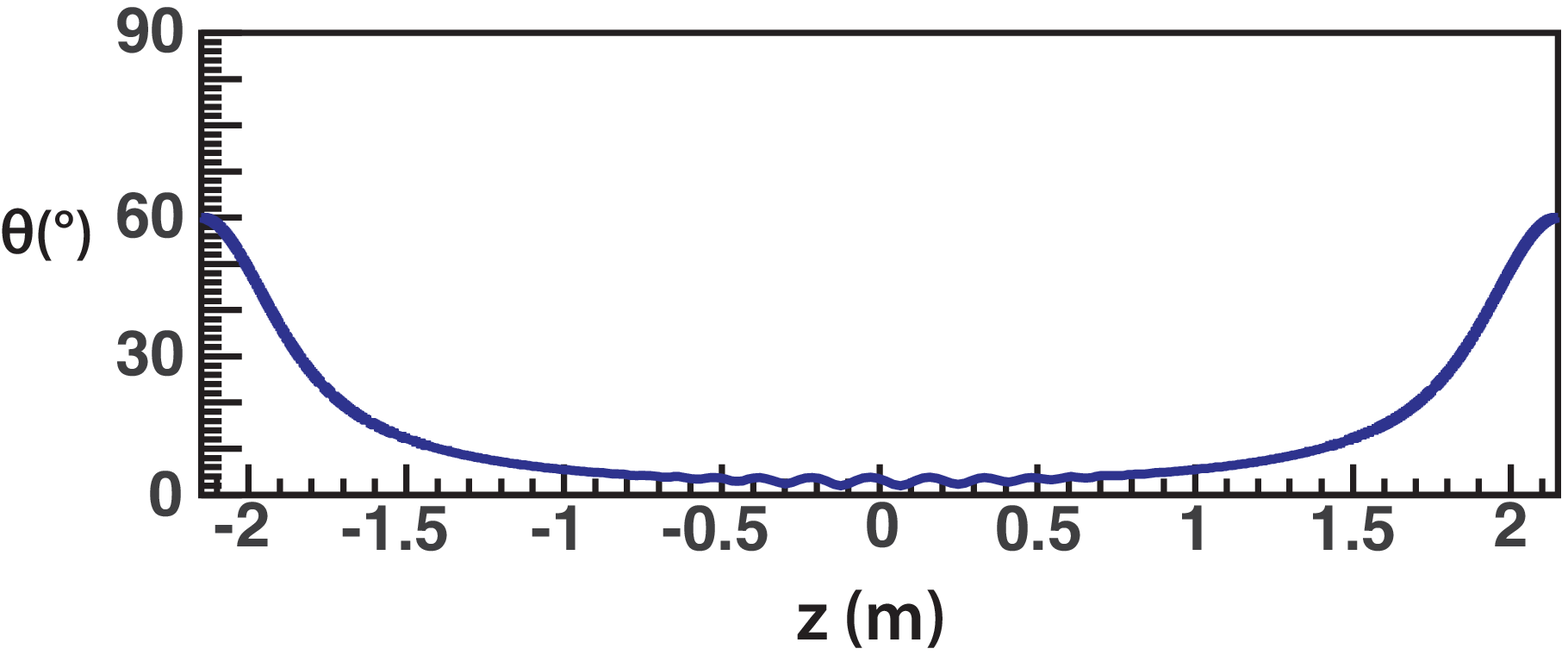}}
\caption{Evolution of the polar angle $\theta$ for a transmitted electron.
Parameters as in Fig. \ref{fig::adi_regeneration}.} \label{fig::angle_adiabatic} \end{minipage} \hfill \begin{minipage}[t]{8cm} \setlength{\fboxsep}{0mm} \centerline{\includegraphics[height=3.3cm]{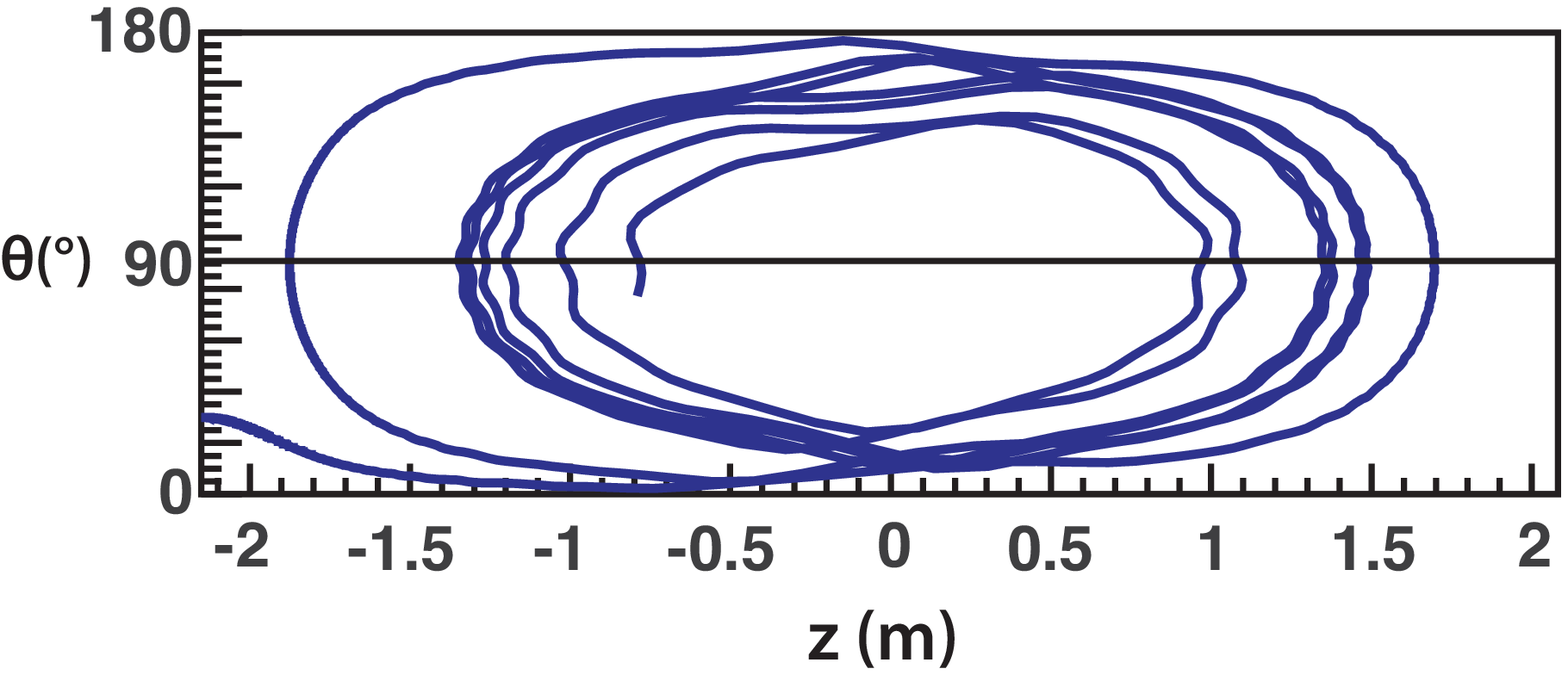}}
\caption{Evolution of the polar angle $\theta$ for a trapped electron.
Parameters as in Fig. \ref{fig::adi_change_trapped}.} \label{fig::trapped_angle} \end{minipage} \end{figure}

The local behaviour of the quantity $\gamma\mu$ was considered as adiabatic invariant in Section \ref{sec::mac_e_principle}.
There, the difference between the electron energy $E$ and the filter energy $qU_{\textrm{f}}$ was assumed to be small (some tens of eV).
As one can see in Fig. \ref{fig::adi_regeneration},  $\gamma\mu$ has an oscillational behavior, due to the superposition of the cyclotron motion and the azimuthal magnetron motion; the oscillation period in Fig. \ref{fig::adi_regeneration} is equal to the electron cyclotron period.
The amplitude of the oscillation
depends on the electron
surplus energy $E_{\textrm{sur}}=E-qU_{\textrm{PS}}$:
With $E_{\textrm{sur}}\approx 0.3$~keV in the PS, the relative fluctuation $\delta  \gamma\mu/\gamma\mu$ inside the PS is order of $10^{-2}$ (for smaller starting polar angle $\theta$ the fluctuation is somewhat larger), but with  $E_{\textrm{sur}}$ of a few keV this fluctuation is much larger, order of 1 (Fig. \ref{fig::adi_regeneration}).

How is it then possible that $\gamma\mu$ regains its starting value, in spite of its large oscillations in the small magnetic field region?
The explanation is the following \cite{Cle69,Den93,Nor63}: the real adiabatic invariant $I_{\textrm{ad}}$, which is constant in the adiabatic
 approximation throughout the whole trajectory, is not $\gamma\mu$, but a complicated function of the higher field derivatives. Inside 
homogeneous field regions the field derivatives are small, and there $  I_{\textrm{ad}} \approx \gamma\mu$. When an electron moves 
from one homogeneous field region to another, and if the adiabatic invariant $I_{\textrm{ad}}$ is constant,
$\gamma\mu$ can regain the starting value with high accuracy, although between these 
two regions, where the field gradients are large, $\gamma\mu $ can have large oscillations.
Using a human analogy, we can say that 
the electron seems to have 'memory', remembering 
its initial value of $\gamma\mu$ \cite{Nor63}. If the electron motion is not adiabatic (in case of high energy or small magnetic field),
$I_{\textrm{ad}}$ is not constant, so in this case the electron will not gain back its original $\gamma\mu $ value in the
second homogeneous field region; then, the electron has no memory.

 In order to illustrate this phenomenon, we simulated electrons starting in the entry-side magnet 
of the PS. At the KATRIN field $B_{\textrm{sol}}=4.5\;\textrm{T}$ (Fig. \ref{fig::katrin_overview} and \ref{fig::prespec_setup}), electrons are
 transmitted through the PS, and $\gamma\mu$ regains its original value (Fig. 
\ref{fig::adi_regeneration}). If we choose a deliberately low field $B_{\textrm{sol}}=0.57\;\textrm{T}$ in simulations, electrons can be
 magnetically trapped and $\gamma\mu$ becomes chaotic (Fig. 
\ref{fig::adi_change_trapped}). If $\gamma\mu$ regains its original value, the angle $\theta$ is approximately determined by Eq. 
\ref{eq::angle_formula}. Let us consider an electron which enters the PS in the entry-side magnet with $\theta<90^{\circ}$ and for which 
the polar angle $\theta$ is determined by Eq. \ref{eq::angle_formula}.
In this case, the electron will never acquire $\theta=90^{\circ}$ anywhere in the PS. Thus, it cannot be magnetically reflected (Fig. 
\ref{fig::angle_adiabatic}). On the other hand, if Eq. \ref{eq::angle_formula} is violated as shown in Fig. 
\ref{fig::adi_change_trapped},
  the reflection angle $\theta=90^{\circ}$ can be reached over and over again (Fig. \ref{fig::trapped_angle}), the electron is magnetically
 trapped inside the PS. It is not transmitted, and so it is not detected.

We mention that, at the edge of outer field lines of the flux tube, with $B_{\rm sol}$=4.5 T and zero PS potential,  the 18 keV electrons make inside the PS about 2 degrees magnetron motional rotation around the beam axis. This has to be taken into account for precise imaging investigations of the KATRIN experiment (see Ref. \cite{Pic92b} for experimental examples of much larger magnetron motional rotation).

The background as a function of the PS potential will be investigated experimentally when the whole KATRIN system is finished. Similarly, the MS transmission function and thus the adiabaticity of the electrons in PS and MS can be experimentally investigated rather precisely, as function of PS potential, if PS and MS are connected together (by shooting E-GUN electrons through them).
The experiments presented in our paper, using only the PS, are sensitive only to large deviations from adiabaticity (small non-adiabaticity effects do not cause any transmission losses in the case of large surplus energies).

\section{Conclusion and Outlook}
\label{sec::conclusion_and_outlook}

Our investigations show that the PS filter energy $qU_{\textrm{PS}}$ can be reduced by several keV without any loss of transmission, making it possible to diminish the Penning trap between PS and MS.
The actual value of the PS filter energy $qU_{\textrm{PS}}$ to minimize KATRIN's background has to be determined experimentally with the full KATRIN setup. It is possible that the final KATRIN setup will operate 
with a mixture of reduced PS filter energy and other, active measures removing stored electrons from the Penning trap between PS and MS. Sweeping a wire through the trapping volume \cite{Bec10}
has proven to be an efficient means to empty the trap. It will take about $t_{\textrm{wire}}=1$~s for the wire to sweep across the magnetic flux tube imaged on the detector. 
In order to scan the tritium $\beta$-spectrum, the MS retarding voltage $U_{\textrm{MS}}$ will be changes every few minutes. The sweeps will be performed during these voltage changes to avoid a loss
of measurement time. Yet, the corresponding $10^6$ motion cycles are a very large number for an UHV compatible device.

For KATRIN, the transmission function of the PS has to be known with permille accuracy \cite{Ang04}. Our investigations indicate that this precision is only possible if the influence of 
multiple electron reflection at the electron source is suppressed. This can be achieved with a) a stable, pulsed electron source at the entry of the PS, b) the PS and MS in their final tandem configuration (cf. Fig. \ref{fig::katrin_overview}) and c) KATRIN's final detector having a time resolution of about 100 ns. 

A novel, angular selective pulsed UV laser photoelectron source, which can produce pulses as short as 40 ns with a repetition rate of up to 10 kHz, is currently being built for
KATRIN on the basis of \cite{Val09,Hug10,Val11}. In the experiment we propose, the MS will be operated so that the electrons retain only a few eV surplus energy in the MS and are therefore guaranteed to be 
transmitted if the MS performs as it should. The filter energy of the PS will be varied, just as in our experiment. In this configuration, multiple (up to about 14) reflections between the detector and the main-spectrometer potential will cease after less than 20 $\mu$s. The fraction of backscattered electrons, which retain enough energy to pass the MS filter potential in backwards (towards electron source) direction and could possibly get lost at the electron source, is negligible in this
configuration. Therefore, the analysis of this experiment will be much simpler than in our case.

\section{Acknowledgement}
\label{sec::acknowledgement}
The authors would like to thank the Kassiopeia development team at KIT and MIT for their valuable work, making the simulations and therefore also the analysis leading to this publication possible.
The pre-spectrometer team, Stefan Kern, Patrick Kr\"amer, Alan Kumb,  Luisa Sch\"afer and Hans Skacel could always be counted on and were a great help in 
keeping the experiment operational. Matthias Prall would particularly like to thank Rainer Gumbsheimer for his hospitality at KIT, for numerous discussions, for his advice and support.
We thank E. Otten for carefully reading the manuscript and for his valuable comments.
This project was supported by the German Ministry for Education and Research (BMBF) under contract number 05A08PM1.

\section*{References}
\bibliography{referenzen_prespec}
\bibliographystyle{iopart-num}

\end{document}